\documentclass[a4paper,english,aps,prl,twocolumn]{revtex4}
\usepackage[T1]{fontenc}
\usepackage[latin1]{inputenc}
\usepackage{verbatim}
\usepackage{amsmath}
\usepackage{graphicx}
\usepackage{amssymb}

\makeatletter
\usepackage{babel}
\makeatother
\begin{document}

\title{Shannon Information Capacity of Discrete Synapses}

\author{Adam B. Barrett and M.C.W. van Rossum}

\address{Institute for Adaptive and Neural Computation\\
University of Edinburgh, 5 Forrest Hill\\
Edinburgh EH1 2QL, UK}

\begin{abstract}
There is evidence that biological synapses have only a fixed number
of discrete weight states. Memory storage with such synapses behaves
quite differently from synapses with unbounded, continuous weights
as old memories are automatically overwritten by new memories. We
calculate the storage capacity of discrete, bounded synapses in terms
of Shannon information. For optimal learning rules, we investigate
how information storage depends on the number of synapses, the number
of synaptic states and the coding sparseness. 
\end{abstract}
\maketitle
Memory in biological neural systems is believed to be stored in the
synaptic weights. Various computational models of such memory systems
have been constructed in order to study their properties and to explore
potential hardware implementations. Storage capacity, and optimal
learning rules have been studied both for single-layer associative
networks \citep{willshaw69,dayan91}, studied here, and for auto-associative
networks \citep{hopfield82,meunier_in_arbib}. Commonly, a synaptic
weight in such models is represented by an unbounded continuous real
number. However, more realistically, synaptic weights have values
between some biophysical bounds. Furthermore, synapses might be restricted
to occupy a limited number of synaptic states. Consistent with this,
some experiments show that, physiologically, synaptic weight changes
occur in steps \citep{petersen98,O'Connor2005}. In contrast to networks
with continuous, unbounded synapses, in networks with discrete, bounded
synapses old memories are overwritten by new ones, in other words,
the memory trace decays \citep{Nadal1986,Parisi1986,Amit1994}. 

It is common to use the signal-to-noise ratio (SNR) to quantify memory
storage \citep{Fusi2007,dayan91}. When weights are unbounded, each
stored pattern has the same SNR, and storage can simply be defined
as the maximum number of patterns for which the SNR is larger than
some fixed, minimum value. For discrete, bounded synapses performance
must be characterized by two quantities: the initial SNR, and its
decay rate. Altering the learning rules typically results in either
1) a decrease in initial SNR but a slower decay of the SNR (i.e.~an
increase in memory lifetime) \citep{Fusi2007}, or 2) an increase
in initial SNR but a decrease in memory lifetime. Optimization of
the learning rule is ambivalent because an arbitrary trade-off must
be made between these two effects. 

The conflict between optimizing learning and optimizing forgetting
can be resolved by analyzing the capacity of synapses in terms of
Shannon information. Here we describe a framework for calculating
the information capacity of bounded, discrete synapses, and use it
to find optimal learning rules. We model a single neuron, and investigate
how the information capacity depends on the number of synapses and
the number of synaptic states, both for dense and sparse coding. 

We consider a single neuron which has $n$ inputs. At each time step
it stores a $n$-dimensional binary pattern with independent entries
$x^{a}$, $a=1\ldots n$. The sparsity $p$ corresponds to the fraction
of entries in $x$ that cause strengthening of the synapse. It is
optimal to set the low state equal to $-p$, and the high state to
$q=:(1-p)$, so that the probability density for inputs is given by
$P(x)=q\delta(x+p)+p\delta(x-q)$ and $\langle x\rangle=0$. The case
$p=\frac{1}{2}$ we term dense, furthermore, we assume that $p\leq\frac{1}{2}$,
as the case $p\geq\frac{1}{2}$ is fully analogous. Although biological
coding is believed to be sparse, we briefly note that in biology the
relation between $p$ and coding \emph{}sparseness is likely very
complicated.

Each synapse occupies one of $W$ states. The corresponding
values of the weight are assumed to be equidistantly spaced around
zero and are written as a $W-$dimensional vector, i.e.~for a 3-state
synapse $\boldsymbol{w}=\{-1,0,1\}$, while for a 4-state synapse
$\boldsymbol{w}=\{-2,-1,1,2\}$. In numerical analysis we have sometimes
seen an increase in information by varying the values of the weight
states, however this increase was always small. Note, that $\boldsymbol{w}$
is very different from the definition of a {}``weight vector'' commonly
used in network models.

The learning paradigm we consider is the following: during the learning
phase a pattern is presented each time step, and the synapses are
updated in an unsupervised manner. The learning algorithm is \emph{on-line,}
i.e.~the synapses can only be updated when the pattern is presented.
As bounded, discrete synapses store new memories at the expense of
overwriting old ones, we can assume that sufficient patterns are stored
such that the earliest pattern has almost completely decayed and the
distribution of the synaptic weights has reached an equilibrium.

After learning, the neuron is tested on learned and novel patterns.
Presentation of a learned pattern will yield an output which is on
average larger than that for a novel pattern. The presentation of
a novel, random pattern $\{ x_{u}^{a}\}$ leads to a signal $h_{u}=\sum_{a}x_{u}^{a}\omega_{a}$,
where the weights are $\omega_{a}$, $a=1,\ldots,n$. As this novel
pattern will be uncorrelated to the weight, it has mean $\left\langle h_{u}\right\rangle =n\left\langle x\right\rangle \left\langle w\right\rangle =0$,
and variance\begin{equation}
\left\langle \Delta h_{u}^{2}\right\rangle =n\left[\left\langle x^{2}w^{2}\right\rangle -\left\langle x\right\rangle ^{2}\left\langle w\right\rangle ^{2}\right]=npq\left\langle w^{2}\right\rangle \,,\label{eq:noise}\end{equation}
where $\left\langle w\right\rangle =\boldsymbol{w}.\boldsymbol{\pi}^{\infty}$,
$\left\langle w^{2}\right\rangle =\sum_{i=1}^{W}w_{i}^{2}\pi{}_{i}^{\infty}$,
and $\mathbf{\boldsymbol{\pi}^{\infty}}$ is the equilibrium distribution
of weights.

Because the synapses are independent, and the performance is characterized
statistically, we can use Markov transition matrices to define the
learning \citep{fusi2002hsd,Fusi2007}. If in the learning phase an
input is high (low), the synapse is updated according to the matrix
$M^{+}$ ($M^{-}$). Thus, the distribution of potentiated weights
immediately after a high input is $\boldsymbol{\pi}^{+}(t=0)=M^{+}\boldsymbol{\pi}^{\infty}$.
As subsequent, uncorrelated, patterns are learned, this signal decays
according to $\boldsymbol{\pi}^{+}(t)=M^{t}\boldsymbol{\pi}^{+}(t=0)$,
where $M=:pM^{+}+qM^{-}$ is the expected update matrix at each time-step.
The equilibrium distribution $\boldsymbol{\pi}^{\infty}$ is identical
to the eigenvector of $M$ with eigenvalue one. The mean signal for
learned patterns is\begin{equation}
\left\langle h_{\ell}\right\rangle (t)=npq\boldsymbol{w}^{T}M^{t}(M^{+}-M^{-})\boldsymbol{\pi}^{\infty}\,.\label{eq:signal}\end{equation}
This signal decays such that the synapses contain most information
on more recent patterns. The decay is typically exponential, with
a time constant equal to the sub-dominant eigenvalue of $M$. 

When tested with an equal mix of learned and unlearned patterns, the
mutual information in the neuron's output about whether a single pattern
is learned or not is\begin{eqnarray}
I & = & \sum_{h,s=\{ u,l\}}P(s)P(h|s)\log_{2}\frac{P(h|s)}{P(h)}\label{eq:Iformula}\\
 & = & \tfrac{1}{2}\sum_{h}P_{\ell}(h)\log_{2}\tfrac{2P_{\ell}(h)}{P_{\ell}(h)+P_{u}(h)}+P_{u}(h)\log_{2}\tfrac{2P_{u}(h)}{P_{\ell}(h)+P_{u}(h)}\nonumber \end{eqnarray}
where $P_{\ell}$ ($P_{u}$) denotes the distribution of the output
of the neuron to learned (unlearned) patterns. If the two output distributions
are perfectly separated, the learned pattern contributes one bit of
information, whilst total overlap implies zero information storage.
As the patterns are independent, the total information is the sum
of the information over all patterns presented during learning.

Unfortunately, the full distributions of $h$ are complicated multinomials.
Furthermore, it would be challenging for a biological readout to distinguish
between two aribitrary distributions. Instead we impose a threshold
between two possible responses, which could, say, correspond to the
neuron firing or not. If the number of synapses is large, we can approximate
the distribution of $h$ with a Gaussian and the information reduces
to a function of the SNR\begin{equation}
I(t)=1+r(t)\log_{2}r(t)+[1-r(t)]\log_{2}[1-r(t)]\,,\label{eq:I-snr}\end{equation}
where $r(t)=\frac{1}{2}\mathrm{erfc}(\sqrt{\mathrm{SNR(t)}/8})$,
and the SNR is defined as\begin{equation}
\mathrm{SNR}(t)=\frac{{2\left(\left\langle h_{\ell}\right\rangle (t)-\left\langle h_{u}\right\rangle \right)}^{2}}{\left\langle \Delta h_{\ell}^{2}\right\rangle (t)+\left\langle \Delta h_{u}^{2}\right\rangle }\,.\label{eq:SNRfull}\end{equation}
In the numerical simulations we use Eqs.\ (\ref{eq:I-snr}) and (\ref{eq:SNRfull}),
but for the analytical expressions we assume the same variance of
the output for learned and unlearned patterns, $\left\langle \Delta h_{\ell}^{2}\right\rangle (t)\approx\left\langle \Delta h_{u}^{2}\right\rangle $.
Importantly, the information (\ref{eq:I-snr}) is a saturating function
of the SNR, and for very high SNR, the information is approximately
one bit. Meanwhile for small SNR, the information is linear in the
SNR, $I\approx\mathrm{SNR}/(4\pi\ln2)$.

%
{}The total information per synapse is obtained by summing together
the information of all patterns and dividing by the number of synapses,
thus $I_{\mathrm{S}}=:{\frac{1}{n}\sum}_{t=0}^{\infty}I[\mathrm{SNR}(t)]$.
In cases in which the initial SNR is very low \begin{equation}
I_{\mathrm{S}}\approx\frac{1}{4\pi n\ln2}\sum_{t=0}^{\infty}\mathrm{SNR}(t)\,.\label{eq:Ilow}\end{equation}
In the opposite limit, when the initial SNR is very high, recent patterns
contribute one bit. We approximate as if all patterns with more than
1/2 bit actually contribute one bit, whilst all patterns with less
information contribute nothing. In this limit the information thus
equals the number of patterns with more than 1/2 bit of information\begin{equation}
I_{\mathrm{S}}=\frac{t_{c}}{n}\,,\label{eq:Ihigh}\end{equation}
where $t_{c}$ is implicitly defined as $I(t_{c})=1/2$. 

The storage capacity depends on the $W\times W$ learning matrices
$M^{+}$ and $M^{-}$. To find the maximal storage capacity we need
to optimize these matrices, and this optimization will in general
depend on the sparseness, the number of synapses, and the number of
states per synapse. Because these are Markov transition matrices,
their columns need sum to one, leaving $W(W-1)$ free variables per
matrix. For dense patterns ($p=1/2$) one can impose additional symmetry
$(M^{+})_{ij}=(M^{-})_{W-i,W-j}$.

In the case of binary synapses we write \begin{equation}
M^{+}=\left(\begin{array}{cc}
1-f_{+} & 0\\
f_{+} & 1\end{array}\right),\;\; M^{-}=\left(\begin{array}{cc}
1 & f_{-}\\
0 & 1-f_{-}\end{array}\right)\,.\label{eq:}\end{equation}
We first consider the limit of few synapses, for which the initial
SNR is low, and use (\ref{eq:Ilow}) to compute the information. (
We keep $np>1$ and $n\gtrsim10$ to ensure that there are sufficient
distinct patterns to learn.) We find \begin{equation}
I_{\mathrm{S}}=\frac{pq}{\pi\ln2}\frac{f_{+}^{2}f_{-}^{2}}{(pf_{+}+qf_{-})^{3}}\frac{1}{2-pf_{+}-qf_{-}}\,.\label{eq:I_W2}\end{equation}
The values for $f_{+}$ and $f_{-}$ that yield maximal information
depend on the value of the density $p$. For $0.11<p<0.89$, one has
$f_{+}=f_{-}=1$, which gives equilibrium weight distribution $\boldsymbol{\pi}_{}^{\infty}=(q,p)^{T}$,
and \begin{equation}
I_{\mathrm{S}}=\frac{pq}{\pi\ln2}\,.\label{eq:I_W2_largep}\end{equation}
In this case the synapse is modified at every time-step and only retains
the most recently presented pattern; the information stored on one
pattern drops to zero as soon as the next pattern is learned. 

For sparser patterns $p<0.11$ a second solution to Eq.~(\ref{eq:I_W2})
is optimal, for which $f_{+}=1$, $f_{-}\approx2p$. I.e.~potentiation
occurs for every high input, but given a low input, depression only
occurs stochasticly with a probability $2p$. As a result, forgetting
is not instantaneous and the SNR decays exponentially with time constant
$\tau=1/(6p)$. The associated weight distribution $\boldsymbol{\pi}^{\infty}\approx(2/3,1/3)^{T}$,
which is interesting to compare to experiments in which about 80\%
of the synapses were found to be in the low state \citep{O'Connor2005}.
The information per synapse is\begin{equation}
I_{\mathrm{S}}=\frac{1}{\pi\ln2}(\frac{2}{27}+\frac{p}{9})\,.\label{eq:I_w2_smallp}\end{equation}
There are two important observations to be made from Eqs.\ (\ref{eq:I_W2_largep}-\ref{eq:I_w2_smallp}):
1) the information remains finite at low $p$, 2) as long as the approximation
is valid, each additional synapse contributes to the information.

We next consider the limit of many synapses, for which the initial
SNR is high. With Eq.~(\ref{eq:Ihigh}) we find\begin{equation}
I_{\mathrm{S}}=\frac{1}{2\ln\left[1-f_{+}p-f_{-}q\right]}\ln\left[\frac{s}{4npq}\frac{(f_{+}p+f_{-}q)^{2}}{f_{+}^{2}f_{-}^{2}}\right]\label{eq:}\end{equation}
where $s\approx6.02$ is defined as the value of the SNR which corresponds
to 1/2 bit of information. The optimal learning parameters are in
this limit $f_{+}=e\sqrt{sq/pn}$ and $f_{-}=e\sqrt{sp/qn}$, leading
to an equilibrium weight distribution $\boldsymbol{\pi}^{\infty}=(1/2,1/2)^{T}.$
In this regime the learning is stochastic, with the probability for
potentiation/depression decreasing as the number of synapses increases.
The corresponding information is\begin{equation}
I_{\mathrm{S}}=\frac{1}{2e\sqrt{spqn}}\approx\frac{0.075}{\sqrt{pqn}}\,.\label{eq:binaryinf}\end{equation}
Hence, as $n$ becomes large, adding extra synapses no longer leads
to substantial improvement in the information storage capacity. The
memory decay time constant is $\tau=\sqrt{n}/(4e\sqrt{spq})$.

\begin{figure}
\includegraphics[width=8cm]{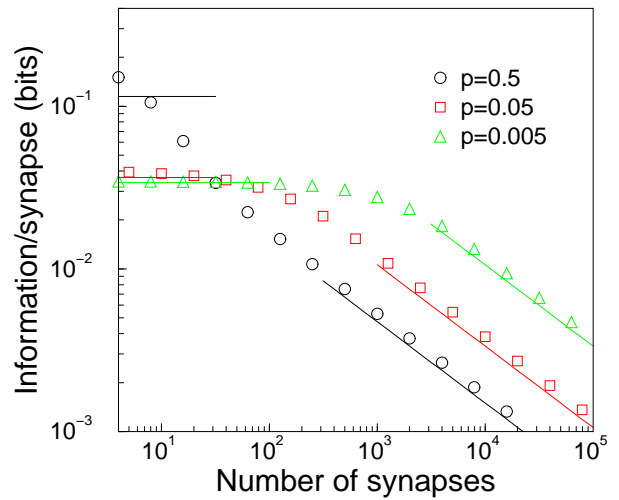}

\caption{Information capacity of binary synapses. Information storage capacity
per synapse versus the number of synaptic inputs, for dense (p=0.5),
sparse (p=0.05) and very sparse (p=0.005) coding. Lines show analytic
results, whilst points show numerical results. For small number of
synapses, each additional synapse contributes to the information.
However, for many synapses, the information per synapse decreases
as $1/\sqrt{n}$.\label{fig:binaryinf}}
\end{figure}

To verify the above results, we carried out a numerical optimization
of learning matrices. We find there is a smooth interpolation between
the two limiting cases, and for a given sparseness there is a critical
number of synapses beyond which the addition of further synaptic inputs
does not substantially improve information storage capacity. This
occurs when the initial SNR becomes of order 1. For dense patterns,
this occurs for just a few synapses, whilst for sparse patterns this
number is proportional to $p^{-1}$, see Fig.~\ref{fig:binaryinf}.

It is interesting to compare the storage capacity found here with
that of a Willshaw net \citep{willshaw69}, which also involves binary
synapses. In the Willshaw model, prior to learning, all synapses occupy
the low state, whilst the learning process consists solely of potentiation
of certain synapses. This means that as more patterns are presented,
more synapses move to the up state, and eventually all memories are
lost. However, when only a finite, optimal number of patterns are
presented, such synapses perform well. When the task is to successfully
reproduce an output pattern associated with each input pattern, the
maximum capacity is 0.69 bits/synapse \citep{willshaw69,Brunel1994}.
When measured for a binary response, as in the framework of this paper,
this value is reduced. In the limit of few synapses, and sparse patterns,
the storage capacity is approximately 0.11 bits/synapse, which is
several times higher than the storage we obtain here. However, as
the number of synapses increases, the storage capacity becomes proportional
to $n^{-1}$, a faster decay than the $n^{-1/2}$ we find for our
case.

Next, we examine whether storage capacity increases as the number
of synaptic states increases. Even under small or large $n$ approximations,
the information is in general a complicated function of the learning
parameters, due to the complexity of the invariant eigenvector $\boldsymbol{\pi}^{\infty}$
of the general Markov matrix $M$. Thus, the optimal learning must
be found numerically by explicitly varying all matrix elements. For
large $n$ we find that the optimal transfer matrix is band diagonal,
with the only transitions allowed being one-step potentiation or depression.
Moreover, we find that for a fixed number of synaptic states, the
(optimized) information storage capacity behaves similarly to that
for binary synapses. In the dense ($p=1/2)$ case, in the limit of
many synapses, the optimal learning rule takes the simple form\begin{equation}
M=\frac{1}{2}\left(\begin{array}{ccccccc}
2-f & 1 & 0\\
f & 0 & 1\\
0 & 1 & 0\\
0 & 0 & 1\\
 &  &  & \ddots\\
 &  &  &  & 0 & 1 & 0\\
 &  &  &  & 1 & 0 & f\\
 &  &  &  & 0 & 1 & 2-f\end{array}\right)\,,\label{eq:optrule}\end{equation}
with $f=e\sqrt{s/n}$ . The equilibrium weight distribution is, somewhat
surprisingly, peaked at both ends, and is low and flat in the middle,
$\boldsymbol{\pi}^{\infty}\propto(1,f,f,\ldots,f,1)^{T}$. The information
is \begin{equation}
I_{\mathrm{S}}=\frac{W-1}{2fn}\ln\frac{f^{2}n}{s}=\frac{W-1}{e\sqrt{sn}}\,,\label{eq:Igeneral1}\end{equation}
and the corresponding time constant for the SNR is given by $\tau=(W-1)\sqrt{n/s}/(2e)$.
Validity of these results requires $fW$ to be small, to enable series
expansion in $f$. Hence, we find that information grows linearly
with the number of synaptic states, provided $W/\sqrt{n}\ll1/(e\sqrt{s})$. 

There appears to be no simple optimal transfer matrix in the sparse
case, even in the large $n$ limit. However, a formula for the storage
capacity which fits well with numerical results and is consistent
with equations \eqref{eq:binaryinf} and \eqref{eq:Igeneral1} is\begin{equation}
I_{\mathrm{S}}=\frac{W-1}{2e\sqrt{spqn}}\,.\label{eq:Ibest}\end{equation}
Assuming that this formula, as for the binary synapse, is the leading
term in a series expansion in the two parameters $f_{+}=e\sqrt{sq/pn}$
and $f_{-}=e\sqrt{sp/qn},$ and that we need $Wf_{+}$ and $Wf_{-}$
small for it to be accurate, then its validity condition is $W\sqrt{q/np}\ll1/(e\sqrt{s})$.

\begin{figure}
\includegraphics[width=8cm]{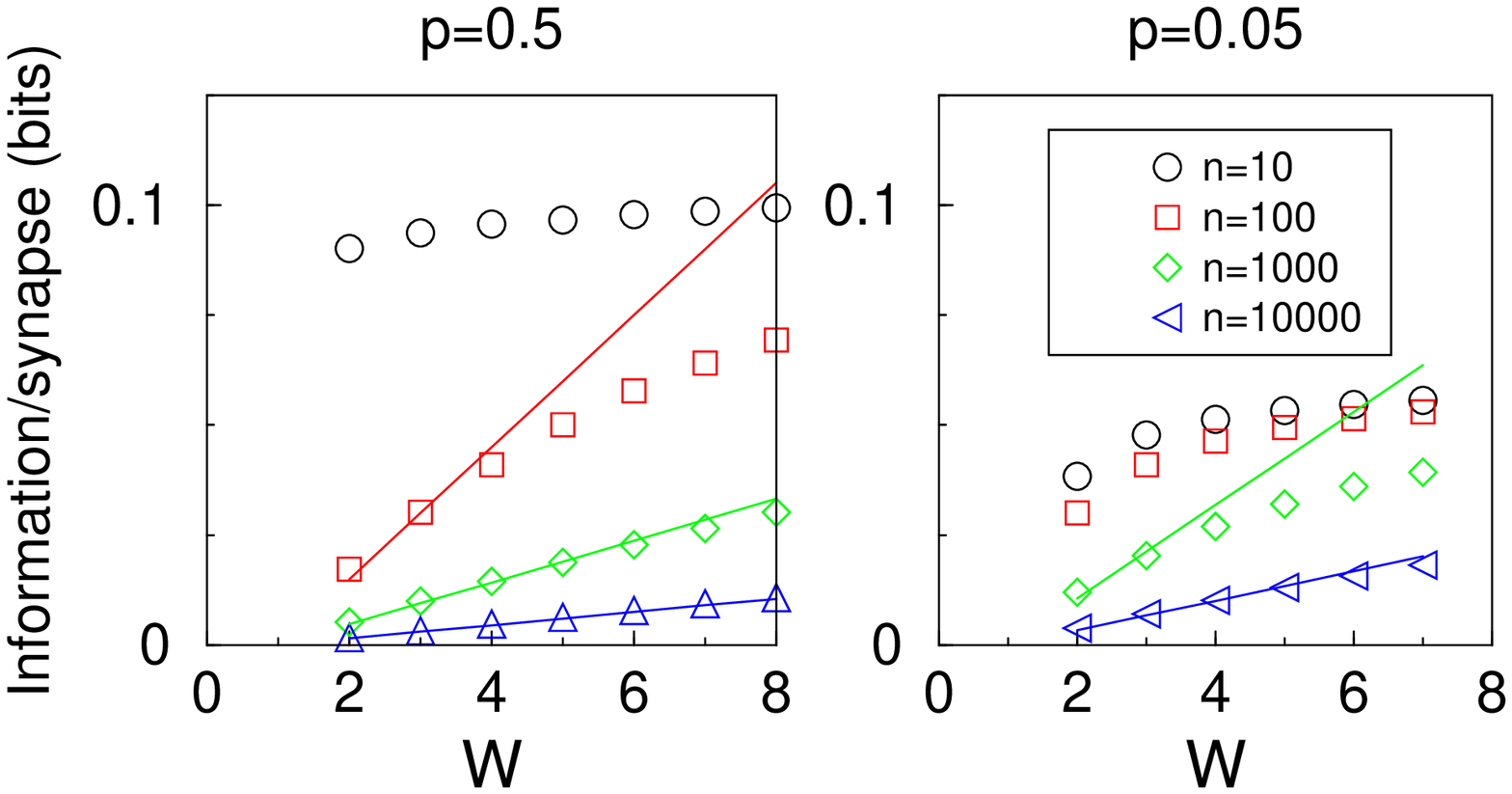}

\caption{Information capacity of multi-state synapses. Information storage
capacity per synapse versus the number $W$of synaptic states, for
dense ($p=0.5$) and sparse ($p=0.05$) coding. Lines show analytic
results (when available), whilst points show numerical results. For
small number of synaptic states, storage capacity is increasing, but
for larger numbers saturates. \label{fig:IvsW}}
\end{figure}

Numerical results agree with the equations above, and are illustrated
together with the analytic results in Fig.~\ref{fig:IvsW}. Thus,
for fixed number of synapses, storage initially grows linearly with
$W$. However, as $W$ becomes larger, capacity saturates and becomes
independent of $W$. This behavior is consistent with that of a number
of different (sub-optimal) learning rules studied in Ref.~\citep{Fusi2007}.
These learning rules had the property that the product of the initial
SNR and the time-constant $\tau$ of SNR decay is independent of $W$
(see Table 1 in \citep{Fusi2007} for this remarkable identity, noting
that the SNR there equals its square root here). For large $W$, or
equivalently small $n$, the initial SNR is small, and hence the information
$I\sim\sum_{t}\mathrm{SNR(0)}\exp(-t/\tau)\sim\mathrm{SNR}(0)\tau$
is independent of $W$, as observed here, Fig.~\ref{fig:IvsW}.

It is interesting to note that even unbounded synapses store only
a limited amount of information. In the framework of this paper, the
optimal local learning rule for unbounded synapses (optimized as in
\citep{dayan91}) yields $\mathrm{SNR}=n/m$, where $m$ is the number
of patterns. This corresponds to storing 0.11 bits/synapse in the
case that $m\gg n\gg1$.

Finally we study, in the large $n$ limit, the performance of a simple
{}``hard-bound'' learning rule, i.e.~a learning rule with uniform
equilibrium weight distribution. Under this rule, whose SNR dynamics
were previously studied in Ref.~\citep{Fusi2007}, a positive (negative)
input gives one step potentiation (depression) with probability $f_{+}$
($f_{-}$). For $W\geq4$ the optimal probabilities satisfy $f_{_{+}}p=f_{-}q\approx e\sqrt{s}W\sqrt{(W+1)/(W-1)}/(2\sqrt{3n})$
and the information storage capacity is\begin{equation}
I_{\mathrm{S}}\approx\sqrt{\tfrac{3(W-1)}{spqn(W+1)}}\frac{1}{eW}\left[1-\cos(\tfrac{\pi}{W-1})\right]^{-1}\!\!\approx0.053\times\frac{W}{\sqrt{pqn}}\,.\end{equation}
Here the latter approximation is for large $W.$ This sub-optimal
learning rule gives an information capacity of the same functional
form as the optimal learning rule, but performs only 70\% as well.%
{}

Given that simple stochastic learning performs almost as well as the
optimal learning rule, one may wonder how well a simple deterministic
learning rule performs in comparison. For certain potentiation and
depression, $f_{+}=f_{-}=1$ , one finds \begin{equation}
I_{\mathrm{S}}=\frac{W^{2}}{\pi^{2}n}\ln\left(\frac{12n}{W^{2}s}\right)\,.\end{equation}
Although this grows faster with $W$, the $1/n$ behavior means this
performs much worse than optimal stochastic learning rules. The memory
decay time is in this case $\tau=W^{2}/\pi^{2}.$

In summary, learning using bounded, discrete weights can be analyzed
using the Shannon information. There are two regimes: 1) when the
number of synapses is small, the information per synapse is constant
and approximately independent of the number of synaptic states, 2)
when the number of synapses is large, the capacity per synapse decreases
as $1/\sqrt{n}$. Furthermore, we find that in the second regime,
the optimal transition matrices are band diagonal. In particular,
in the dense case ($p=1/2$), the matrix Eq.~(\ref{eq:optrule})
is optimal. The critical $n$ that separates these regimes is dependent
on the sparseness and the number of weight states. Although there
are currently no good biological estimates for either the number of
weight states, or the sparsity, these results might indicate that
the number of synapses is limited to prevent sub-optimal storage.
When increasing the number of synaptic states, we find that for small
numbers the information grows linearly, while for larger numbers it
levels off and reaches a saturation point. 

%
{}

\begin{acknowledgments}
This work was supported by the HFSP. We thank Henning Sprekeler, Peter
Latham, Jesus Cortes, Guy Billings and Robert Urbanczik for discussion.
\end{acknowledgments}

\bibliographystyle{apsrev}

\begin{thebibliography}{12}
\expandafter\ifx\csname natexlab\endcsname\relax\def\natexlab#1{#1}\fi
\expandafter\ifx\csname bibnamefont\endcsname\relax
  \def\bibnamefont#1{#1}\fi
\expandafter\ifx\csname bibfnamefont\endcsname\relax
  \def\bibfnamefont#1{#1}\fi
\expandafter\ifx\csname citenamefont\endcsname\relax
  \def\citenamefont#1{#1}\fi
\expandafter\ifx\csname url\endcsname\relax
  \def\url#1{\texttt{#1}}\fi
\expandafter\ifx\csname urlprefix\endcsname\relax\def\urlprefix{URL }\fi
\providecommand{\bibinfo}[2]{#2}
\providecommand{\eprint}[2][]{\url{#2}}

\bibitem[{\citenamefont{Willshaw et~al.}(1969)\citenamefont{Willshaw, Buneman,
  and Longuet-Higgins}}]{willshaw69}
\bibinfo{author}{\bibfnamefont{D.~J.} \bibnamefont{Willshaw}},
  \bibinfo{author}{\bibfnamefont{O.~P.} \bibnamefont{Buneman}},
  \bibnamefont{and} \bibinfo{author}{\bibfnamefont{H.~C.}
  \bibnamefont{Longuet-Higgins}}, \bibinfo{journal}{Nature}
  \textbf{\bibinfo{volume}{222}}, \bibinfo{pages}{960} (\bibinfo{year}{1969}).

\bibitem[{\citenamefont{Dayan and Willshaw}(1991)}]{dayan91}
\bibinfo{author}{\bibfnamefont{P.}~\bibnamefont{Dayan}} \bibnamefont{and}
  \bibinfo{author}{\bibfnamefont{D.~J.} \bibnamefont{Willshaw}},
  \bibinfo{journal}{Biol. Cybern.} \textbf{\bibinfo{volume}{65}},
  \bibinfo{pages}{253} (\bibinfo{year}{1991}).

\bibitem[{\citenamefont{Hopfield}(1982)}]{hopfield82}
\bibinfo{author}{\bibfnamefont{J.~J.} \bibnamefont{Hopfield}},
  \bibinfo{journal}{Proc. Natl. Acad. Sci.} \textbf{\bibinfo{volume}{79}},
  \bibinfo{pages}{2554} (\bibinfo{year}{1982}).

\bibitem[{\citenamefont{Meunier and Nadal}(1995)}]{meunier_in_arbib}
\bibinfo{author}{\bibfnamefont{C.}~\bibnamefont{Meunier}} \bibnamefont{and}
  \bibinfo{author}{\bibfnamefont{J.-P.} \bibnamefont{Nadal}}, in
  \emph{\bibinfo{booktitle}{The handbook of Brain theory, 1st edition}}, edited
  by \bibinfo{editor}{\bibfnamefont{M.~A.} \bibnamefont{Arbib}}
  (\bibinfo{publisher}{MIT press}, \bibinfo{address}{Cambridge, MA},
  \bibinfo{year}{1995}).

\bibitem[{\citenamefont{Petersen et~al.}(1998)\citenamefont{Petersen, Malenka,
  Nicoll, and Hopfield}}]{petersen98}
\bibinfo{author}{\bibfnamefont{C.~C.~H.} \bibnamefont{Petersen}},
  \bibinfo{author}{\bibfnamefont{R.~C.} \bibnamefont{Malenka}},
  \bibinfo{author}{\bibfnamefont{R.~A.} \bibnamefont{Nicoll}},
  \bibnamefont{and} \bibinfo{author}{\bibfnamefont{J.~J.}
  \bibnamefont{Hopfield}}, \bibinfo{journal}{Proc. Natl. Acad. Sci.}
  \textbf{\bibinfo{volume}{95}}, \bibinfo{pages}{4732} (\bibinfo{year}{1998}).

\bibitem[{\citenamefont{O'Connor et~al.}(2005)\citenamefont{O'Connor,
  Wittenberg, and Wang}}]{O'Connor2005}
\bibinfo{author}{\bibfnamefont{D.~H.} \bibnamefont{O'Connor}},
  \bibinfo{author}{\bibfnamefont{G.~M.} \bibnamefont{Wittenberg}},
  \bibnamefont{and} \bibinfo{author}{\bibfnamefont{S.~S.-H.}
  \bibnamefont{Wang}}, \bibinfo{journal}{Proc Natl Acad Sci U S A}
  \textbf{\bibinfo{volume}{102}}, \bibinfo{pages}{9679} (\bibinfo{year}{2005}).

\bibitem[{\citenamefont{Nadal et~al.}(1986)\citenamefont{Nadal, Toulouse,
  Changeux, and Dehaene}}]{Nadal1986}
\bibinfo{author}{\bibfnamefont{J.}~\bibnamefont{Nadal}},
  \bibinfo{author}{\bibfnamefont{G.}~\bibnamefont{Toulouse}},
  \bibinfo{author}{\bibfnamefont{J.}~\bibnamefont{Changeux}}, \bibnamefont{and}
  \bibinfo{author}{\bibfnamefont{S.}~\bibnamefont{Dehaene}},
  \bibinfo{journal}{Europhysics Letters (EPL)} \textbf{\bibinfo{volume}{1}},
  \bibinfo{pages}{535} (\bibinfo{year}{1986}).

\bibitem[{\citenamefont{Parisi}(1986)}]{Parisi1986}
\bibinfo{author}{\bibfnamefont{G.}~\bibnamefont{Parisi}}, \bibinfo{journal}{J.
  Phys. A: Math. Gen} \textbf{\bibinfo{volume}{19}}, \bibinfo{pages}{L617}
  (\bibinfo{year}{1986}).

\bibitem[{\citenamefont{Amit and Fusi}(1994)}]{Amit1994}
\bibinfo{author}{\bibfnamefont{D.}~\bibnamefont{Amit}} \bibnamefont{and}
  \bibinfo{author}{\bibfnamefont{S.}~\bibnamefont{Fusi}},
  \bibinfo{journal}{Neural Computation} \textbf{\bibinfo{volume}{6}},
  \bibinfo{pages}{957} (\bibinfo{year}{1994}).

\bibitem[{\citenamefont{Fusi and Abbott}(2007)}]{Fusi2007}
\bibinfo{author}{\bibfnamefont{S.}~\bibnamefont{Fusi}} \bibnamefont{and}
  \bibinfo{author}{\bibfnamefont{L.~F.} \bibnamefont{Abbott}},
  \bibinfo{journal}{Nat Neurosci} \textbf{\bibinfo{volume}{10}},
  \bibinfo{pages}{485} (\bibinfo{year}{2007}).

\bibitem[{\citenamefont{Fusi}(2002)}]{fusi2002hsd}
\bibinfo{author}{\bibfnamefont{S.}~\bibnamefont{Fusi}},
  \bibinfo{journal}{Biological Cybernetics} \textbf{\bibinfo{volume}{87}},
  \bibinfo{pages}{459} (\bibinfo{year}{2002}).

\bibitem[{\citenamefont{Brunel}(1994)}]{Brunel1994}
\bibinfo{author}{\bibfnamefont{N.}~\bibnamefont{Brunel}},
  \bibinfo{journal}{Phys. A} \textbf{\bibinfo{volume}{27}},
  \bibinfo{pages}{4783} (\bibinfo{year}{1994}).

\end{thebibliography}

\end{document}